\documentclass{article}
\usepackage{arxiv}
\usepackage[utf8]{inputenc} 
\usepackage[T1]{fontenc}    
\usepackage{hyperref}       
\usepackage{url}            
\usepackage{booktabs}       
\usepackage{amsfonts}       
\usepackage{nicefrac}       
\usepackage{microtype}      
\usepackage{lipsum}		
\usepackage{graphicx}
\usepackage{multirow}
\usepackage{doi}
\usepackage{cite}
\usepackage{amsmath,amssymb,amsfonts}
\usepackage{algorithmic}
\usepackage{graphicx}
\usepackage{textcomp}
\usepackage{xcolor}
\usepackage{marvosym}
\usepackage{stfloats}
\usepackage{booktabs}
\usepackage{threeparttable}
\usepackage{subcaption}
\usepackage{mwe}

\title{AICOM-MP: an AI-based Monkeypox Detector for Resource-Constrained Environments}

\author{Tim~Tianyi~Yang, Tom~Tianze~Yang, Andrew~Liu, Jie~Tang, Na~An, Shaoshan~Liu, and~Xue~Liu}

\date{}

\renewcommand{\headeright}{}
\renewcommand{\undertitle}{}

\hypersetup{
}

\begin{document}
\maketitle

\begin{abstract}
Under the Autonomous Mobile Clinics (AMCs) initiative, we are developing, open sourcing, and standardizing health AI technologies to enable healthcare access in least developed countries (LDCs). We deem AMCs as the next generation of health care delivery platforms, whereas health AI engines are applications on these platforms, similar to how various applications expand the usage scenarios of smart phones. Facing the recent global monkeypox outbreak, in this article, we introduce AICOM-MP, an AI-based monkeypox detector specially aiming for handling images taken from resource-constrained devices. Compared to existing AI-based monkeypox detectors, AICOM-MP has achieved state-of-the-art (SOTA) performance.  We have hosted AICOM-MP as a web service to allow universal access to monkeypox screening technology. We have also open sourced both the source code and the dataset of AICOM-MP to allow health AI professionals to integrate AICOM-MP into their services. Also, through the AICOM-MP project, we have generalized a methodology of developing health AI technologies for AMCs to allow universal access even in resource-constrained environments. 
\end{abstract}


\section{Introduction}

\begin{figure}
\centering
\includegraphics[width=0.8\linewidth]{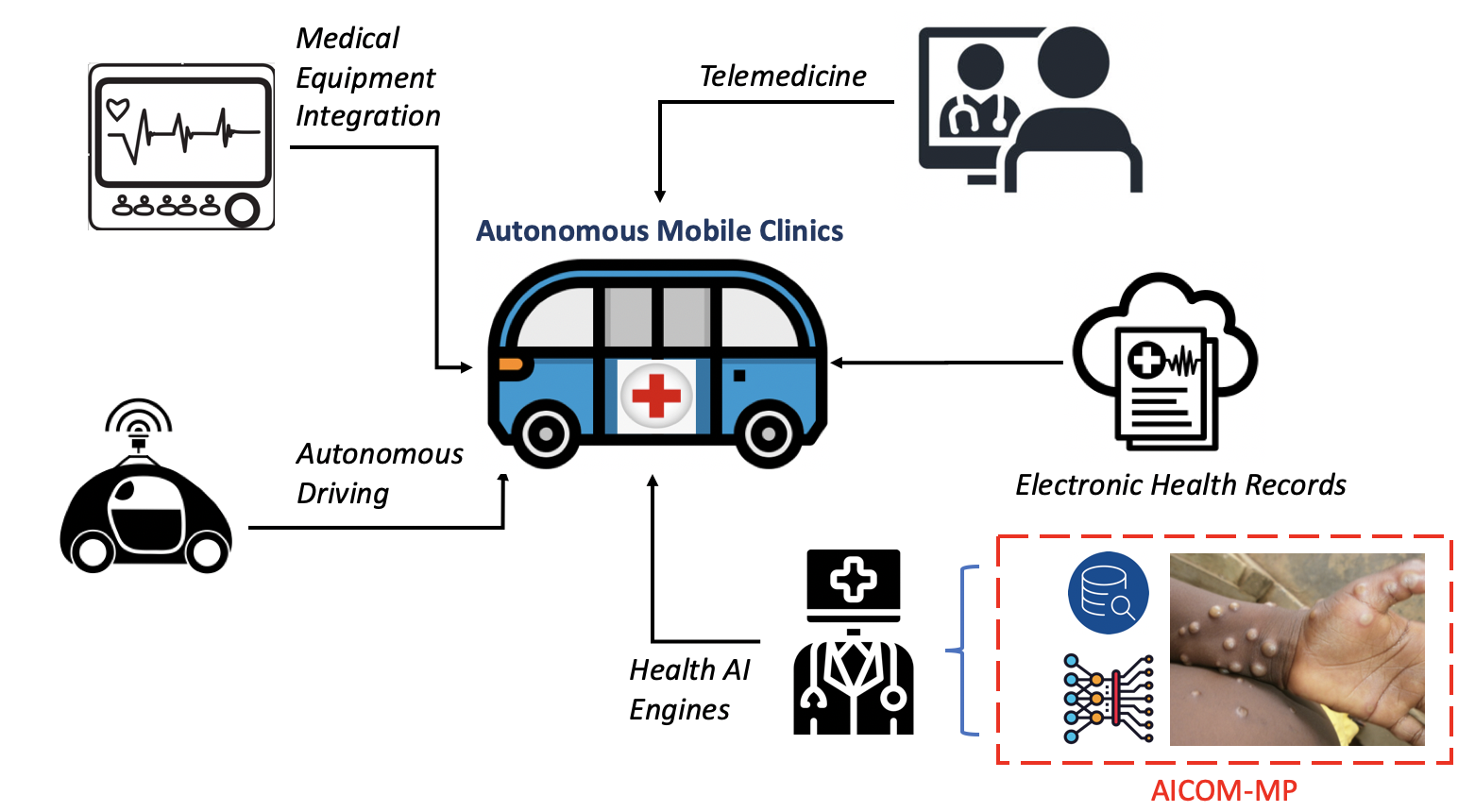}
\caption{\textbf{Enabling technologies for Autonomous Mobile Clinics }}
\label{fig:amc}
\end{figure}

As illustrated in Fig.~\ref{fig:amc}, autonomous mobile clinics are self-driving vehicles equipped with medical diagnostic equipment, telemedicine capability and an artificial intelligence (AI) software that can perform some of the tasks of health professionals, such as disease screening and basic diagnostics. Such clinics have the potential to revolutionize health-care delivery by bringing health-care services to hard-to-reach populations ~\cite{liu2022autonomous}. 

Particularly, AMCs answer to the United Nations Sustainable Development Goal 3 (SDG3) on health, which represents a universal recognition that health is fundamental to human capital and social and economic development ~\cite{sdg}. Despite the progress achieved at the global level, many health systems are not sufficiently prepared to respond to the needs of the rapidly aging population. Inequitable access to healthcare continues to impede progress towards achieving universal health coverage (UHC). AMCs empower the provision of essential health services that’s safe, affordable, and effective, with a special emphasis on the poor, vulnerable and marginalized segments of the population~\cite{wef}.

Because AMCs are autonomous, these clinics can independently navigate to places where health-care services are needed, while their mobility allows health-care access on-site. These clinics are equipped with conversational and visual user interfaces to facilitate their use by all population groups. Particularly, the health AI engines incorporated into these clinics can serve an unlimited number of patients, and only cases that cannot be handled by the software would require the intervention of remote doctors, thus potentially reducing the cost for the human workforce~\cite{liu2022autonomous}.

While AMCs serve as mobility platforms for essential health services provisioning, the health AI engines incorporated onto AMCs determine the reach and coverage of AMCs. In essence, health AI engines are applications on AMCs, similar to how various applications expand the usage scenarios of smart phones. This way, AMCs should be configured with different medical devices and health AI engines as the deployment situation demands ~\cite{liu2022rise}. 

Under the AMCs initiative and facing the recent global monkeypox outbreak, in this article we introduce AICOM-MP, an AI-based monkeypox detector that has achieved SOTA performance and capable of handling images taken from resource constrained devices, such as low-end mobile phones, computing environments often endured by people in LDCs. Currently, AICOM-MP exists in the form of a web service, such that it can be accessed through equipment within AMCs, or through a wide spectrum of mobile devices. Hence, AICOM-MP is required to handle pictures taken by various devices, especially low-resolution images taken on resource-constrained devices. 

Through this project, we have made the following contributions to the field of health AI:
\begin{itemize}
  \item Datasets are key to machine learning model training, we have constructed the AICOM-MP dataset to optimize coverage, diversification, and generalization, and we have open-sourced the dataset for other health AI professional to improve upon (\textbf{Section ~\ref{sec:dataset}}).
  \item We have developed AICOM-MP in the form of web service to allow universal access of monkeypox screening technology. We have open-sourced AICOM-MP source code for other health AI professional to integrate AICOM-MP into their services (\textbf{Section ~\ref{sec:arch}}).
  \item We have demonstrated that AICOM-MP trained on AICOM-MP dataset effectively filters out noises from real-world data and achieves SOTA detection results. Particularly, the restoration unit within the AICOM-MP pipeline effectively handles images taken on low-end devices (\textbf{Section ~\ref{sec:eval}}).
  \item With the methodology introduced in this article, we have generalized a methodology for developing health AI engines to enrich the coverage of AMCs (\textbf{Section ~\ref{sec:diss}}).
\end{itemize}

\section{Monkeypox Symptoms and Screening Methods}
\label{sec:monkeypox}

Monkeypox is a highly contagious skin disease which causes rash-like lesions ~\cite{WHO_monkeypox}. Up to Oct 24 2022, there have been 75,568 global cases in 109 locations, 74,677 of which occurred in locations that did not report any monkeypox cases historically ~\cite{cdc}. The lesion regions can be seen on several body parts including hands, faces, genital areas, and normally turn into scabs before healing. Patients may experience symptoms such as fever, swollen lymph nodes, and can give rise to various medical complications such as sepsis. Note that the rash tends to be more concentrated on the face and extremities rather than on the trunk. It affects the face (in 95\% of cases), and palms of the hands and soles of the feet (in 75\% of cases). 

Clinical diagnoses for Monkeypox can be conducted through a biopsy or a polymerase chain reaction (PCR) laboratory test ~\cite{WHO_monkeypox}. However, these tests are expensive, and LDCs may not be able to afford mass-scale PCR testing.  In contrast, utilizing AI for monkeypox screening is cost efficient and can be deployed in mass-scale in LDCs. AI has shown great potential to carry out productive monkeypox diagnoses. Multiple research groups (~\cite{sitaula2022monkeypox}~\cite{Nafisa2022}) have trained monkeypox classification machine learning models. Moreover, Islam et al ~\cite{islam2022can} and Patel et al  ~\cite{patel2022artificial} demonstrated the prospects of detecting monkeypox through deep learning CV models.   

Different from other existing works, we have developed AICOM-MP according to the following principles: minimization of gender, racial, and age bias; ability to conduct binary classification when the image resolution/quality is low; capacity to produce accurate results irrespective of images' background. Also our model focuses on detection of the rash of the face and extremities to maximize screening accuracy.  In addition, we have carefully constructed a AICOM-MP training dataset to meet the above requirements. The AICOM-MP dataset is also balanced so that the amount of monkeypox images and "others"(i.e. other diseases, healthy, etc.) images are approximately equal. Our approach has achieved SOTA performance for AI-based monkeypox detection.

\section{Monkeypox Dataset for Model Training}
\label{sec:dataset}

Datasets are key to machine learning model training, in this section we review existing monkeypox datasets, summarize the shortcomings of existing datasets, and explain how the AICOM-MP Dataset that we developed addresses these problems. The resulting AICOM-MP training dataset contains 4932 images, and the validation/testing dataset contains 360 images that are split by a ratio of 65/35 (234 validation images and 126 testing images), all of which have equal number of images from the two classes. The AICOM-MP dataset is available at ~\cite{aicomgithub}.

\subsection{Existing Monkeypox Datasets}

We start by performing a comprehensive survey of existing monkeypox datasets as follows:

\begin{itemize}
  \item \textbf{Monkeypox Skin Image Dataset 2022}: Islam et al ~\cite{Monkeypox_Skin_Image_Dataset_2022} populated this Monkeypox dataset through web-scraping  Monkeypox, Chickenpox, Smallpox, Cowpox, and Measles infected skin as well as healthy skin images. This dataset contains 804 original web-scrapped images, including 160 Monkeypox images, 178 chickenpox images, 358 smallpox images, 54 cowpox images, 47 measles images, and 50 healthy-skin images. These images have been sequentially screened by two expert physicians specialized in infectious diseases to validate the supposed infection.
  \item \textbf{Monkeypox-dataset-2022}: Ahsan et al ~\cite{Monkeypox-dataset-2022} generated this dataset also through web-scraping from websites, newspapers, and online portals and publicly shared samples using internet search engines. This dataset contains 171 original web-scrapped images, including 43 monkeypox images, 47 chickenpox images, 17 measles images, and 54 healthy-skin images. Seven augmentation operations have been applied to the web-scrapped images using ImageDataGenerator from Keras.
  \item \textbf{Monkeypox Skin Lesion Dataset}: Ali et al  ~\cite{Nafisa2022} developed this dataset through manual web-scrapping from public case reports, news portals, and websites. Skin lesion images were verified using Google’s Reverse Image Search and cross-referenced with other sources. The repeated, out-of-focus, low-resolution, and low-quality images were discarded using a two-stage screening process. The dataset contains 228 original web-scrapped images, including 102 monkeypox images, and 126 others (i.e., chickenpox,measles) images. 13 augmentation operations were applied to the web-scrapped images.
\end{itemize}   

Despite different approaches of organizing and augmenting image data have been applied in each dataset, shortage of high quality monkeypox lesion images and imbalance of images from torso, race, and age groups remain to be the major problems. Most of the monkeypox images have low resolutions, and some of them are artificially generated. These raise our concern that the machine learning model may treat certain resolution level as a sign of monkeypox images. 




\subsection{AICOM-MP Dataset}

\begin{figure}[htb]
\centering
\includegraphics[width=0.8\linewidth]{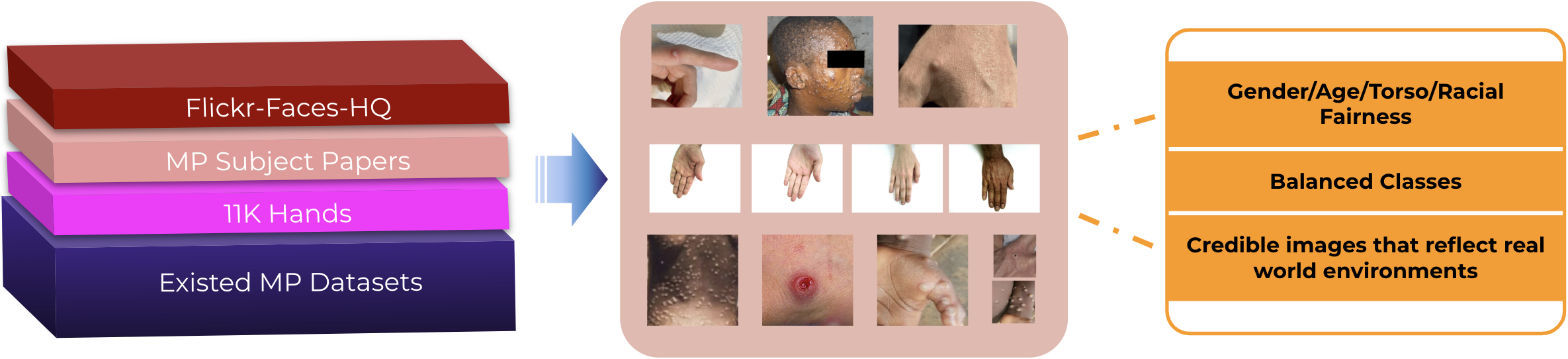}
\caption{\textbf{AICOM-MP Dataset}}
\label{fig:datasetMP}
\end{figure}

As illustrated in Figure \ref{fig:datasetMP}, we utilized ~\cite{Nafisa2022} as the foundation dataset, then we enriched and balanced the dataset through including more unique monkeypox images that exhibit salient monkeypox lesion features, over-sampling the dataset to include equal number of healthy skin images from different torso, gender, racial, and age groups, and using carefully selected techniques to augment images from the minority class (monkeypox) until reaching a balanced state with respect to the majority class (others) among the dataset. To improve the model's generalization ability such that it can accurately map the noisy and transformed images to an accurate categorization, we have trained the AICOM-MP model on the augmented dataset but used the original un-augmented images for validation.


\subsubsection{Data Selection}

On the basis of ~\cite{Nafisa2022}, we have added 18 healthy skin images and 8 monkeypox images from ~\cite{sitaula2022monkeypox}. Additionally, We selected 18 healthy hand and 18 healthy face images from the 11k hands dataset ~\cite{11kHands} and  Flickr-Faces-HQ Dataset~\cite{Karras_2019_CVPR}. We also collected 22 newly-found monkey-pox images from seven published articles whose subject matter is monkeypox disease (~\cite{mccollummonkeypox,kabugamonkeypox,petersenmonkeypox,giromettimonkeypox,brunomonkeypox,minhajmonkeypox,mdmonkeypox}). As a result, the AICOM-MP dataset contains 312 original images, of which 132 images are classified to monkeypox and 180 images are classified into others. We used this dataset as our validation/testing dataset which was split by a ratio of 65/35. To verify how well our trained model performs under real world settings, we generated a new test dataset \textit{COCO\_MP} dataset, which consists of 200 manually selected COCO 2017 dataset~\cite{cocodataset} and 132 original monkeypox images. The diversity of sources of the AICOM-MP dataset ensures the coverage of our trained models.




\subsubsection{Data Augmentation}
 
Data augmentation increases the coverage, diversification, and the generalization ability of training dataset. It helps the model perform better in real-world environments. Augmentations were executed based on image geometric manipulations (rotation, translation, noise injection, color space transformations) and did not incorporate photometric transformations for quality control purposes, as photometricly permutated images makes little sense from a human perspective~\cite{shorten_khoshgoftaar_2019}. To augment the newly appended images, we followed the techniques applied by the precedent dataset~\cite{Nafisa2022}.

\subsubsection{Data Balancing}

Data balancing acts as a tool against impartial favor towards the majority class(es) during models' internal decision process. After applying such technique, there are 1818 augmented monkeypox images and 2466 augmented 'others' images which are used as the training dataset. We balanced the training dataset by augmenting 618 monkeypox images using the \textit{ImageGenerator} function from Tensorflow Keras. The aforementioned validation/testing dataset was also balanced by augmenting 48 monkeypox images using the same function.

\section{AICOM-MP Monkeypox Detector Architecture}
\label{sec:arch}

In this section, we delve into the AICOM-MP architecture. Note that, we have implemented a web service to host AICOM-MP, which is able to efficiently process images taken on a wide spectrum of mobile devices, such that people around the world can access and utilize the web service for monkeypox screening. Particularly, we have optimized AICOM-MP to process images taken on resource‑limited devices, such that people from LDCs can utilize the AICOM-MP technology.  The AICOM-MP web service can be accessed through~\cite{aicomWeb}.  In addition, the AICOM-MP source code and dataset are available at ~\cite{aicomgithub} to allow health AI professionals to integrate AICOM-MP into their services.

\begin{figure}[htb]
\centering
\includegraphics[width=0.8\linewidth]{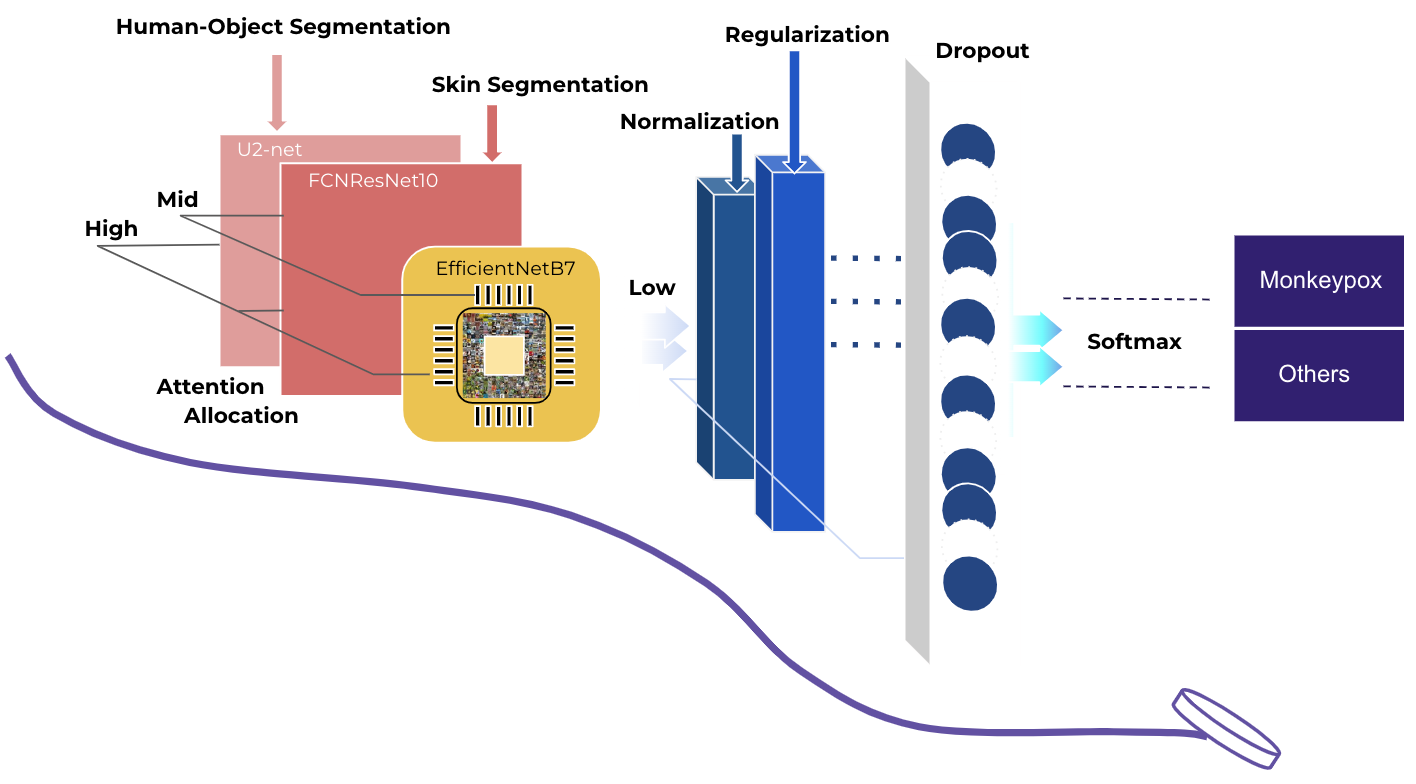}
\caption{\textbf{AICOM-MP Model Architecture}}
\label{fig:arch}
\end{figure}

The AICOM-MP pipeline resembles how professionals diagnose pox symptoms through visual inspections on skin lesions. We model such expertise into the medical AI and perform medical diagnoses based on an expert information retrieval manner. In particular, the image processing step operates similarly to physicians' visual behaviours on allocating their visual attention towards lesion regions. AICOM-MP's model architecture is illustrated in Fig \ref{fig:arch}. 


\subsection{Human-object Segmentation/Background Removal}  

As the first component of the analytical pipeline, this "level-1" vision layer is dedicated to determine the object towards which the vision model should allocate its attention, letting the model know "what" to look for. It plays the role of channeling attention in ViT and constitutes the object selection process. We selected the $U^2$-Net DL model and used its pre-trained weights to carry out object detection and background removal tasks~\cite{U_square_net}. It has a a two-level nested U-structure architecture that promises a computational-efficient yet powerful performance. 

By identifying human objects and removing components of other classes in the image space, $U^2$-net helps increase model's robustness in dealing with images taken in noisy and complex real-world environments, and reduces the chances for the model to misdirect its attention to irrelevant salient objects, such as trees. Additionally, it was noticed that if a human object occupies the majority of space, $U^2$-net would view the object itself as background and removes a large proportion of the object. Based on empirical observation, a blacked-out region that occupies more than 87\% of the original image space would output meaningless results that have no use for medical inferences, and we decided to preserve the image without applying segmentation in that case. 


\subsection{Region-based Skin Detection and Segmentation}  

To further leverage the reduced spatial information from inputted images, this "level-2" vision layer plays the role of enhancing discrimination against non-skin features. It is to constitute the adaptive spatial region selection mechanism of spatial attention in ViT and determines "where" the model should focus to facilitate monkeypox screening processes.       

We selected an open-sourced FCNResNet10 model and used its pre-trained weights to carry out region-based skin detection and segmentation task~\cite{semantic_seg}. The model was trained on 150 images, where skin segmentation annotations were added and which encompass various skin colours and lighting conditions, from the COCO dataset. Consistent with the previous vision layer, a blacked-out mask that covers more than 87\% of the original image space causes significant loss of information that is medically-valuable, and would be excluded from being applied to the inputted images. 

Based on our empirical evaluations, we noticed improvements brought by blocking regions covered by non-skin objects (i.e. rings, sweaters, etc) from the model's inputting field. Following this discovery, we proceeded to testing the necessity of former layer as to whether there is an advantage of combining these two attention mechanisms. The ablation experiments reveal a notable improvements brought by combining the two models. Given consideration of the unforeseeable real-world environments, it is necessary to combine the two attention mechanisms since the former would prevents the latter from being misled and removing valuable regions from the image space. 






\subsection{Deep-learning (DL) based Classification}
 
As the last component of the pipeline, this "level-3" vision layer is designed to allocate the model's eventual attention towards skin lesions across the reduced image space. It constitutes ViT's branch attention and determines "which" the model is looking for. It is assigned with a classification task of detecting monkeypox features embedded among skin image patches. To enable the use of DL model in energy-constraint devices while maintaining robust monkeypox screening performances under resource-limited conditions, we resort to EfficientNetB7 which has an architecture designed for performance on mobile CPU and uses a simple yet highly effective compound coefficient to scale up CNNs in a more structured manner ~\cite{efficient_net}.

In addition, to boost the model's performance in a limited yet domain-specific dataset, we did a transfer learning on the pretrained ImageNet weights~\cite{deng2009imagenet}. Inspired by model training methods described in ~\cite{kaggleB3model}, four additional layers were attached to the EfficientNetB7 model: a batch normalization layer where momentum = 0.99 and $\epsilon$ = 0.001); a dense layer containing a l2 kernel regularizer with l = 0.016, l1 activity regularizer \& l1 bias regularizer with l = 0.006, and relu activation function; a dropout layer whose dropout rate = 0.45; an output layer with softmax activation function to produce binary classification results.

\begin{figure}
\centering
\includegraphics[width=0.8\linewidth]{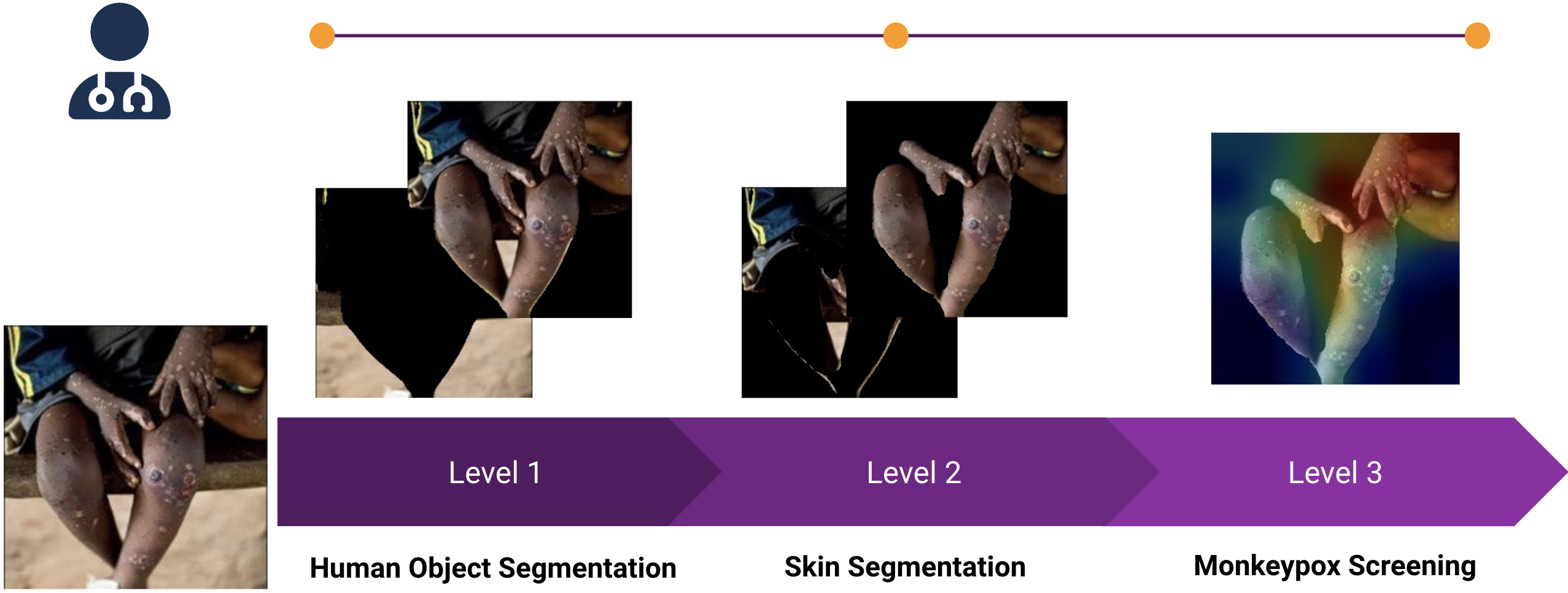}
\caption{\textbf{AICOM-MP Workflow}}
\label{fig:pipeline}
\end{figure}

The end-to-end processing pipeline of AICOM-MP is illustrated in Fig \ref{fig:pipeline}. During the implementation process, we tuned the parameters as follows: each image was resized into 224*224 dimensions; adaptive momentum estimation (Adam) was used as the optimizer; batch size was set at 48; momentum was set at 0.99; and the initial learning rate was set at 0.001. To prevent over-fitting, we utilise the learning rate decay over each epoch coupled with the kernel and activity regularizers. 


\subsection{Resolution Restoration}

One of AICOM-MP's objectives is to handle pictures taken on low-end phones, these pictures usually have low image resolution. To achieve this, we have implemented an resolution restoration unit in the AICOM-MP pipeline. But to avoid altering the perception of monkeypox characteristics by the machine learning model, we use images of original resolution (without enhancing) when training the model. With this enhancement, we have observed a considerable improvement on resolution of monkeypox images. An ablation study that investigated the effectiveness of this component is given in section ~\ref{sec:eval}.

\section{Evaluation}
\label{sec:eval}

In this section, we perform a comprehensive study to evaluate AICOM-MP.  We compare AICOM-MP to existing models on various datasets as well as real-world data to demonstrate that AICOM-MP is the SOTA in monkeypox screening. 

A key contribution of the current work is to make monkeypox screening available in resource-restrained environments, which is essential to enable healthcare access for people from LDCs. To achieve this goal, we only considered families of EfficientNet as potential candidates for DL models and conducted experiments solely on them. We intentionally did not consider models that are computationally demanding, such as ResNet. Through a few stages of optimization, we have managed to achieve high detection accuracy. 

Specifically on the detection accuracy results, Ali et al have presented a performance comparison of different deep learning models(VGG 16, ResNet50, InceptionV3) and an ensemble model. The four models respectively achieved detection accuracy ranging from 74\% to 90\% on the binary monkeypox classification dataset proposed by Ali et al ~\cite{Nafisa2022}. As a comparison, AICOM-MP has achieved 100\% accuracy on Ali et al's binary monkeypox classification tasks. 



To verify that AICOM-MP is not overfitting to achieve perfect scores, we evaluated AICOM-MP on our experimental dataset consisted of 6124 images, in which 2596 images belong to the monkeypox class and 3528 images belong to the other class, augmented using different parameter settings than the training augmentation settings. Note that, our focus is not solely on improving the classification accuracy of our model, but also on improving the existing datasets by taking into account of fairness issues (torso/gender/racial/age/bias), discovering related methods to balance datasets, and designing an analytical pipeline suitable for medical practices and real-world challenges. \textbf{Conclusion 1: As shown in Table \ref{tab:perf}, AICOM-MP outperforms other proposed methods, and has achieved SOTA results on various monkeypox datasets.}


\begin{table}[htb]
\caption{\textbf{Model Performance Statistics on Various Datasets}}
\begin{tabular}{|p{0.5\linewidth}|p{0.13\linewidth}|p{0.145\linewidth}|p{0.13\linewidth}|}
 \hline
     \multicolumn{4}{|c|}{Performance Comparison on the Experimental Dataset} \\
 
\hline
Model                                                             & Weighted-Pre. & Weighted-Recall & Weighted-F1    \\ \hline
Adjusted EfficientNetB3 trained on ~\cite{Nafisa2022} dataset                       & 0.8673            & 0.8601          & 0.8608                    \\ \hline
AICOM-MP trained on ~\cite{Nafisa2022} dataset                       & 0.9102            & 0.8827          & 0.8844                    \\ \hline
AICOM-MP trained on ~\cite{Nafisa2022} dataset + healthy skin images & 0.9295            & 0.9136          & 0.9147                    \\ \hline
AICOM-MP trained on AICOM-MP dataset                             & \textbf{0.9650}   & \textbf{0.9634} & \textbf{0.9635}  \\ \hline
\end{tabular}
\label{tab:perf}
\end{table}

\begin{table}[htb]
\caption{\textbf{Ablation over the COCO\_MP Dataset using Proposed Architecture}}
\begin{tabular}{|cccccc|}
\hline
\multicolumn{6}{|c|}{Ablation Study Results}                                                                                                                                                                                                                                                                                                                                                                                                                                                                                                                                          \\ \hline
\multicolumn{1}{|p{0.11\textwidth}|}{\begin{tabular}[c]{@{}c@{}}AICOM-MP\\ (~\cite{Nafisa2022} dataset)\end{tabular}} & \multicolumn{1}{p{0.18\textwidth}|}{\begin{tabular}[c]{@{}c@{}}AICOM-MP\\(AICOM-MP dataset)\end{tabular}} & \multicolumn{1}{p{0.09\textwidth}|}{\begin{tabular}[c]{@{}c@{}}Restoration\end{tabular}} & \multicolumn{1}{c|}{\begin{tabular}[c]{@{}c@{}}Background Removal\end{tabular}} & \multicolumn{1}{c|}{\begin{tabular}[c]{@{}c@{}}Skin Segmentation\end{tabular}} &  \begin{tabular}[c]{@{}c@{}}
Accuracy (\%)\end{tabular} \\ \hline
\multicolumn{1}{|c|}{$\bullet$}                                                        & \multicolumn{1}{c|}{\textbf{}}                                                              & \multicolumn{1}{c|}{\textbf{}}                                                    & \multicolumn{1}{c|}{\textbf{}}                                                    & \multicolumn{1}{c|}{\textbf{}}                                                   &  40.66                                                       \\ \hline
\multicolumn{1}{|c|}{}                                                         & \multicolumn{1}{c|}{$\bullet$}                                                                      & \multicolumn{1}{c|}{}                                                             & \multicolumn{1}{c|}{}                                                             & \multicolumn{1}{c|}{}                                                            &  88.55                                                       \\ \hline
\multicolumn{1}{|c|}{}                                                         & \multicolumn{1}{c|}{$\bullet$}                                                                      & \multicolumn{1}{c|}{$\bullet$}                                                            & \multicolumn{1}{c|}{}                                                             & \multicolumn{1}{c|}{}                                                            &  91.57                                                        \\ \hline
\multicolumn{1}{|c|}{}                                                         & \multicolumn{1}{c|}{$\bullet$}                                                                      & \multicolumn{1}{c|}{}                                                             & \multicolumn{1}{c|}{$\bullet$}                                                            & \multicolumn{1}{c|}{}                                                            &  95.18                                                        \\ \hline
\multicolumn{1}{|c|}{}                                                         & \multicolumn{1}{c|}{$\bullet$}                                                                      & \multicolumn{1}{c|}{}                                                             & \multicolumn{1}{c|}{}                                                             & \multicolumn{1}{c|}{$\bullet$}                                                           &  95.48                                                        \\ \hline
\multicolumn{1}{|c|}{}                                                         & \multicolumn{1}{c|}{$\bullet$}                                                                      & \multicolumn{1}{c|}{$\bullet$}                                                            & \multicolumn{1}{c|}{$\bullet$}                                                            & \multicolumn{1}{c|}{}                                                            & 96.08                                                        \\ \hline
\multicolumn{1}{|c|}{}                                                         & \multicolumn{1}{c|}{$\bullet$}                                                                      & \multicolumn{1}{c|}{}                                                             & \multicolumn{1}{c|}{$\bullet$}                                                            & \multicolumn{1}{c|}{$\bullet$}                                                           &  \textbf{96.99}                                                        \\ \hline
\multicolumn{1}{|c|}{}                                                         & \multicolumn{1}{c|}{$\bullet$}                                                                      & \multicolumn{1}{c|}{$\bullet$}                                                            & \multicolumn{1}{c|}{}                                                             & \multicolumn{1}{c|}{$\bullet$}                                                           &  95.18                                                        \\ \hline
\multicolumn{1}{|c|}{}                                                         & \multicolumn{1}{c|}{$\bullet$}                                                                      & \multicolumn{1}{c|}{$\bullet$}                                                            & \multicolumn{1}{c|}{$\bullet$}                                                            & \multicolumn{1}{c|}{$\bullet$}                                                           &  \textbf{96.34}                                                        \\ \hline
\end{tabular}
\label{tab:ablation}
\end{table}

Next we evaluated AICOM-MP on real-world data, obtained from COCO\_MP dataset as described in section \ref{sec:dataset}.  From table \ref{tab:perf}, we see that not only the selection of DL models is of the utmost importance, but the quality of training datasets greatly influences the models' performance. Specifically, when we switched from using EfficientNetB3 to using EfficientNetB7 while keeping ~\cite{Nafisa2022} as their common training dataset, we noticed an increase of at least 2.26\% and at most 4.29\% across the models' precision, recall, and F1 score; when we switched from using ~\cite{Nafisa2022} to using AICOM-MP dataset to train a EfficientNetB7 model, the model's accuracy, recall, and F1 score all increased by at least 5.48\% and at most 8.07\%. We also observe such effect in table \ref{tab:ablation}, where we demonstrated that the EfficientNetB7 trained using AICOM-MP dataset achieved an accuracy of 88.55\% on COCO\_MP dataset whereas the EfficientNetB7 trained using ~\cite{Nafisa2022} had an accuracy of 40.66\%. In comparison, AICOM-MP has achieved the highest generalizability in terms of not giving misdiagnoses when handling real-world images, and highest capability of capturing subtle and salient characteristics of monkeypox lesion among noisy environments. \textbf{Conclusion 2: AICOM-MP has achieved SOTA monkeypox detection results on real-world data.}

Finally, we analyze how each technique in the AICOM-MP pipeline improves detection performance, and the results have been summarized in table ~\ref{tab:ablation}.  Comparing to the original AICOM-MP(\cite{Nafisa2022} dataset) model, AICOM-MP (AICOM-MP dataset) model itself increases the accuracy by more than two folds, suggesting not only the accurate representation of the medical information embedded within the dataset, but also its extensive generalization ability to real-world environments. Thus, AICOM-MP (AICOM-MP dataset) exhibited high monkeypox screening performances and it 
constitutes level-3 vision layer. Additionally, using $U^{2}$-Net to suppress background noises and enhance foreground human object(s), the accuracy achieved by level-3 vision layer was increased by 6.63\%, demonstrating a rectifying effect $U^{2}$-Net has on level-3 vision layer. In short, $U^{2}$-Net has proven to be useful in improving level-3 vision layer's monkeypox screening performance and it constitutes level-1 vision layer. Nonetheless, many features from human gadgets, such as sweaters, may appear as indications of monkeypox infections, thereafter inducing incorrect diagnoses. We resorted to a skin segmentation algorithm to construct the level-2 vision layer to prune non-skin features, and successfully increasing the model's accuracy by 6.93\%. We obtained the most optimized performance by further combining level-1, level-2, and level-3 vision layers. The new architecture minimizes the effect that real-world distractions have on the level-3 vision layer as it boosts the model's accuracy to 96.99\%, demonstrating the effectiveness of superposing level-2 vision layer on level-1 vision layer and the efficacy of our designed attention allocation pipeline in medical images processing. \textbf{Conclusion 3: the AICOM-MP pipeline trained on AICOM-MP dataset effectively filters out noises from real-world data and achieves SOTA detection results}.   

In addition, we analyze the effectiveness of the proposed restoration-unit using COCO\_MP dataset. The ablation results shown in the second and third rows of table \ref{tab:ablation} indicate an increase in classification accuracy by 3.02\% when the restoration-unit was used before passing into AICOM-MP. \textbf{Conclusion 4: the restoration unit within the AICOM-MP pipeline effectively handles images taken on low-end devices}.

\section{Discussion}
\label{sec:diss}

Health AI engines are applications on AMCs, this article presents the first case of "application" development on and beyond AMCs. Particularly, starting with monkeypox, we have realized that many existing health AI engines fail to handle a wide spectrum of data, and especially fail to perform on images taken from resource-constrained devices. In this section, we summarize the learning from the AICOM-MP project, and generalize a methodology that can be applied to other health AI "applications" beyond monkeypox, such as measles, chickenpox, and eczema.

\begin{itemize}
  \item \textbf{Robust Dataset}: A robust dataset that has wide coverage, diversification, and generalization is essential to improve detection accuracy of a health AI engine. In the case of AICOM-MP, We have developed a dataset through data selection to improve coverage, data augmentation to improve diversification, and data balancing to improve generalization.  
  \item \textbf{Attention Model}: Through this project, we have identified that a model that resembles how medical professionals diagnose diseases using the attention mechanism performs the best. The image processing step operates similarly to physicians' visual behaviours on allocating their visual attention towards lesion regions. Images are then processed hierarchically through three stages of stratification of visual information, and only the medically-indicative features are kept for further inferences. This attention-based method can be generalized to other health AI engines as well.
  \item \textbf{Compute Optimization}: Compute optimization is imperative to ensure efficiency and scalability of health AI engines so that these engines can serve unlimited patients. In AICOM-MP, we have optimized both data processing and handling images from resource-constrained devices. For data processing, we have incorporated a compressor component in our web service. We have also utilized an image restoration module to improve detection accuracy on pictures taken from low-end mobile devices. 
\end{itemize}

By combining the above techniques, efficient and effective health AI engines can be developed. However, when the health AI engines are still deployed as web services, connectivity is imperative. As a next step, we focus on maximizing the usage of AICOM-MP, especially for people in LDCs. According to the International Telecommunication Union (ITU) 2021 status report, 76\% of the population of LDCs is covered by a mobile broadband signal and only 25\% are using the Internet~\cite{ldc_network}. Despite having a mobile device, most people living in LDCs do not have access to any internet services. Therefore, merely open-sourcing AICOM-MP is insufficient, it is critical that AICOM-MP has the capability to carry out offline diagnoses. To address this problem, we are currently migrating AICOM-MP to low-end mobile devices for offline computing, especially targeting mobile devices most widely used in LDCs.

\section{Conclusion}
\label{sec:concl}
This article is the first to explore the development of health AI applications for AMCs, which are mobility platforms for essential health services provisioning, especially for people with limited health care access. We deem AMCs as the next generation of health care delivery platforms, whereas health AI engines are applications on these platforms, similar to how various applications expand the usage scenarios of smart phones. We chose monkeypox as our starting point as the world is currently under a global monkeypox outbreak, and the technology developed can be immediate help. Not only did we achieve SOTA monkeypox detection results, we have hosted the technology on a web service to allow universal access, we have open sourced the code and dataset for health AI professionals to integrate into their services. Most importantly, the methodology developed in this article can be generalized to many other diseases, leading to the development of many health AI engines for various AMCs usage scenarios. 

\bibliographystyle{unsrt}
\bibliography{references}
\end{document}